# A New Paradigm for Improved Image Steganography by using Adaptive Number of Dominant Discrete Cosine Transform Coefficients

Laeeq Aslam Sandhu[4*], Ebrahim Shahzad[1], Fatima Yaqoob[3], Sharjeel Abid Butt[1], Wasim Khan[1], I. M. Qureshi[2]

[1] Department of Electrical Engineering, International Islamic University, Islamabad, Pakistan
[2] Department of Electrical Engineering, Air University Islamabad, Pakistan.
[3] University of Sargodha.
[4] Department of Automation, Central South University, Changsha, China
Corresponding Author Email: laeeq.aslam.100@gmail.com

**ABSTRACT**

Image steganography camouflages secret messages in images by tampering image contents. There is a natural desire for hiding maximum secret information with the least possible distortions in the host image. This requires an algorithm that intelligently optimizes the capacity keeping the required imperceptibility of the image. This paper presents an image steganography scheme that preserves an adaptively chosen block of dominant coefficients from each Discrete Cosine Transform coefficients, whereas the rest of the coefficients are replaced with normalized secret image pixel values. Secret image pixel value are normalized in an adaptively chosen range. Embedding such kind of normalized data in adaptively chosen non-square L- shaped blocks utilize maximum embedding space available in each block that consequently results in maximizing payload capacity, while maintaining the image quality. This scheme achieved payload capacity up to 21.5 bit per pixel (bpp), while maintaining image quality of 38.24 dB peak signal to noise ratio.

## 1. INTRODUCTION

Greek word "Steganography " is used for an art of camouflaged writing. It is a combination of two words, "Steganos" mean "covered" and "Graphical" mean "writing"[1], [2]. Invisible inks such as "lemon water ink" were frequently used in second world war as mentioned in [3]. Microdots were used by the Germans for the very first time to hide secret messages[3], [4]. In today's communication confidentiality is a evergreen challenge. Encryption schemes and steganography algorithms have been designed to comply this challenge. Encryption algorithms such as RSA (Rivest, Shamir, and Adelman) algorithm [5] and DES (Data Encryption Standard) algorithm [6] were in commercial use. Encryption chippers such as RSA and DES usually scramble information data to make it incomprehensible in other words dubious and attains spotlight of unintended users. Consequently, there is always a chance of decryption. On the other hand Steganography camouflages the secret information existence in a host medium. In literature host medium is also known as cover medium.

Several digital steganography algorithms have been suggested in the last few decades. Substantially most often they share a central idea of inserting secret information in a host medium with the help of an embedding algorithm to produce a stego-output as presented in Fig.1. various steganography algorithms include audio, text,video,network and image steganography. These types are categorized on the basis of their host medium also known as cover..

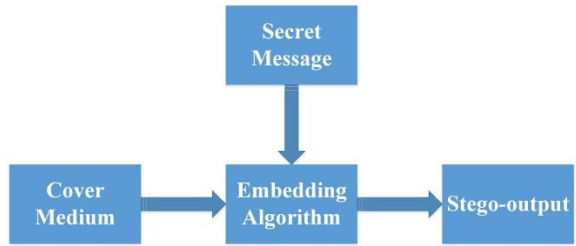

**Fig.1** General Steganography diagram

Furthermore, comparison matrices among various steganography schemes are imperceptibility, payload capacity , and robustness [7].

Image steganography can be classified into two main categories i.e. spatial and frequency domain algorithms. Spatial domain schemes inject secret information precisely in to the pixel intensity values. Authors in [8], [9], proposes two different spatial domain high capacity steganography schemes by utilizing variable size of L.S.B substitution and shared color palette respectively to achieve high capacity while maintaining the resemblance between cover and stego images. Authors of [10] used pixel value differencing (PVD) along with adaptive LSB injection for data hiding in edge area of cover images. In [11], author proposed a data hiding algorithm for the grayscale host images that are compressed using block truncation codes and achieved capacity up to 2 bpp with 33.97 db PSNR, whereas, in frequency domain techniques, image are converted to their frequency domain equivalent by applying Fast Fourier Transform (FFT) [12], Discrete Cosine Transform (DCT)[13]–[18] and Discrete Wavelet Transform (DWT)[19], [20]. Later the high frequency coefficients of the host image (cover image) are removed. As in [17], the author used the standard JPEG quantization matrix with the quality factor 50. Finally, the secret message is embedded in place of high frequency coefficients and inverse transformation is applied to generate stego-image. Among many challenges one is to attain an ability for achieving improved capacity while maintaining quality of a stego image. This helps us to save bandwidth and improve imperceptibility of stego images. Recently proposed scheme such as [17], [21] have shown much better results in term of high capacity data embedding.

In this work, a new discrete cosine transform based image steganography algorithm for high capacity embedding with better imperceptibility is introduced. This algorithm retains only dominant DCT coefficient whereas all other coefficients are replaced with secret image intensity values. Moreover, instead of choosing a square block for secret data embedding as in [17], [21], the proposed scheme hides information in an adaptively chosen non-square $L$ shaped block. Furthermore, existing schemes such as [17], [21] normalize the secret message in a fixed range before embedding, whereas this scheme adaptively chooses the normalization limit for each $8 \times 8$ window. Consequently, this results in improving stego-image quality. Moreover, existing schemes transforms an image into multiple window sizes varying from $8 \times 8$ up to $256 \times 256$ and embed message in all of them to find an optimal window size that gives us the required capacity with best possible imperceptibility. This algorithm divides an image only in $8 \times 8$ non-overlapping blocks and later applies constraints on maximum embedding limit in each block to find an optimal point that gives the required capacity with best possible imperceptibility. Thus this algorithm requires less computations than the schemes discussed in [17], [21]. Therefore, this algorithm has shown improvements in all aspects that includes increasing payload capacity better imperceptibility with fewer computations than existing schemes.

The rest of the paper is organized as follows. In section 2, fundamental mathematical concept required for understanding the proposed scheme is discussed. Section 3 discusses the proposed scheme, whereas results are compared and discussed in section 4. Section 5 discusses computational complexity of the proposed scheme. Toward the end, section 6 concludes the discussion.

## 2. STEGO-IMAGE QUALITY MEASUREMENT

Generally, if the existence of secret message in a cover medium is anticipated, steganography schemes lose their advantages over encryption schemes. Subjective test for image steganography schemes such as discussed in [17] asks people to distinguish between original images and stego-images. If the percentage of realization is below 50% concludes that the secrecy of the algorithms are is intact.

The secret information injected in a host or cover image pretend as noise within a cover image. Beside subjective test, scientific community uses Peak signal to Noise Ratio (PSNR) as a comparison parameter. Author in [8], [14], [15], [22], [23] used PSNR for analyzing quality of stego-image. PSNR is given by

$$PSNR = 20 \, \log_{10} \left( \frac{Max_X}{\sqrt{MSE}} \right) \qquad (1)$$

Where
$$MSE = \frac{1}{M \times N} \sum_{i=0}^{M-1} \sum_{j=0}^{N-1} [F(i,j) - F'(i,j)]^2 \qquad (1.1)$$

Where $F$ and $F'$ represent host and stego-image with dimension $M \times N$. Mean square error ($MSE$) is low if the resemblance between stego and the cover image is high.

## 3. PROPOSED SCHEME

The suggested algorithm for attaining high embedding capacity with better stego-image quality is is discussed in this section. This comprises of an embedding and a retrieving algorithm.

### 3.1 EMBEDDING PROCESS

The embedding process comprises of six steps that takes host and a secret image to generate a stego-image. These steps are explained as,

**STEP 1:** Divide host image into $8 \times 8$ windows. These windows are non-overlapping.

**STEP 2:** Apply 2D-Discrete Cosine Transform on each window using

$$F(u,v) = \frac{2}{N} C(u)C(v) \sum_{x=0}^{N-1} \sum_{y=0}^{N-1} F(x,y) \times \left[\cos(\frac{\pi u(2x+1)}{2N})\right]\left[\cos(\frac{\pi v(2y+1)}{2N})\right] \quad (2)$$

$$C(v), C(u) = \begin{cases} \frac{1}{\sqrt{2}} & if\ u = 0 \\ 1 & otherwise \end{cases} \quad (2.1)$$

Where $F(x,y)$ is the intensity of the host image. The example of 2D- Discrete Cosine Transform coefficients magnitudes is shown in Fig. 2a.

**STEP 3:** Standard JPEG quantization matrix "**Q**" with quality factor 50 is used to quantize the $8 \times 8$ coefficient window $F(u,v)$ using .

$$D(u,v) = round\left\{\frac{F(u,v)}{Q(u,v)}\right\} \quad (3)$$

Fig. 2b shows quantization matrix "**Q**".

**STEP 4:** Find a block of non-zero coefficients having size $K \times K$ in the top left corner of matrix **D**. These are the least number of discrete cosine transform coefficients that are inevitable to hold. Block size is flexibly selected for every window and transmitted as a key with the stego-image. We can restrict the minimum block size. This will help in improving the stego-image quality. This restriction on minimum size is discussed in detail towards the end of this section.

**STEP 5:** The discrete cosine transform coefficients of original host image are replaced back in place of non-zero $K \times K$ block in matrix D. The $3 \times 3$ red boarded area Fig.2c represents non-zero $K \times K$ block. The remaining block in matrix D is a $L$ shape block that is shaded with blue color in Fig. 2c This $L$ shape block is replenished with the normalized values of the secret image. These pixel values are normalised using

$$NSI = \left[\frac{SI}{255}\right] \times NF \quad (4)$$

| 851.00 | 90.77 | 43.38 | 10.01 | 3.75  | 11.16 | 9.36 | 3.81 |
|--------|-------|-------|-------|-------|-------|------|------|
| 99.03  | 5.05  | 25.84 | 23.17 | 14.71 | 12.01 | 1.20 | 4.47 |
| 41.64  | 25.15 | 25.30 | 25.85 | 11.82 | 7.19  | 1.48 | 0.62 |
| 10.24  | 14.34 | 13.41 | 11.91 | 6.95  | 4.09  | 4.10 | 1.82 |
| 6.75   | 15.26 | 9.91  | 5.94  | 4.50  | 2.19  | 2.38 | 2.24 |
| 9.84   | 12.56 | 5.18  | 2.48  | 2.35  | 3.26  | 0.98 | 1.98 |
| 7.10   | 1.39  | 0.98  | 2.26  | 2.40  | 1.21  | 0.19 | 0.64 |
| 3.13   | 3.91  | 1.07  | 3.13  | 2.06  | 0.79  | 0.50 | 0.22 |

(a)

| 16.00 | 11.00 | 10.00 | 16.00 | 24.00  | 40.00  | 51.00  | 61.00  |
|-------|-------|-------|-------|--------|--------|--------|--------|
| 12.00 | 12.00 | 14.00 | 19.00 | 26.00  | 58.00  | 60.00  | 55.00  |
| 14.00 | 13.00 | 16.00 | 24.00 | 40.00  | 57.00  | 69.00  | 56.00  |
| 14.00 | 17.00 | 22.00 | 29.00 | 51.00  | 87.00  | 80.00  | 62.00  |
| 18.00 | 22.00 | 37.00 | 56.00 | 68.00  | 109.00 | 103.00 | 77.00  |
| 24.00 | 35.00 | 55.00 | 64.00 | 81.00  | 104.00 | 113.00 | 92.00  |
| 49.00 | 64.00 | 78.00 | 87.00 | 103.00 | 121.00 | 120.00 | 101.00 |
| 72.00 | 92.00 | 95.00 | 98.00 | 112.00 | 100.00 | 103.00 | 99.00  |

(b)

| 53.00 | 8.00 | 4.00 | 1.00 | 0.00 | 0.00 | 0.00 | 0.00 |
|-------|------|------|------|------|------|------|------|
| 8.00  | 4.00 | 2.00 | 1.00 | 1.00 | 0.00 | 0.00 | 0.00 |
| 3.00  | 2.00 | 2.00 | 1.00 | 0.00 | 0.00 | 0.00 | 0.00 |
| 1.00  | 1.00 | 1.00 | 0.00 | 0.00 | 0.00 | 0.00 | 0.00 |
| 0.00  | 1.00 | 0.00 | 0.00 | 0.00 | 0.00 | 0.00 | 0.00 |
| 0.00  | 0.00 | 0.00 | 0.00 | 0.00 | 0.00 | 0.00 | 0.00 |
| 0.00  | 0.00 | 0.00 | 0.00 | 0.00 | 0.00 | 0.00 | 0.00 |
| 0.00  | 0.00 | 0.00 | 0.00 | 0.00 | 0.00 | 0.00 | 0.00 |

(c)

| 851.00 | 90.77 | 43.38 | | | | | |
|--------|-------|-------|---|---|---|---|---|
| 99.03  | 50.05 | 25.84 | | | | | |
| 41.64  | 25.15 | 25.30 | | | | | |

**Embedded Secret Image Pixel values**

(d)

**Fig. 2** (a) DCT coefficient of 8x8 windows. (b) Standard jpeg quantization matrix Q. (c) Quantized DCT coefficient. (d) DCT coefficient of cover image updated with secret pixel values.

Where $SI$ and $NSI$ are used for secret message and the normalized secret message respectively. The normalization factor $NF$ is selected such that it is the maximum value of the replaced original coefficients. Normalization factor for the example shown in fig 2 is 26. The block size calculated in step 4 and the NF both serve as the secret key. This key is sent along with the stego-images and is used in retrieving process explained in the next sub-section. Cover image updated window is shown in Fig. 2d.

**STEP 6:** Inverse discrete cosine transform of each window is calculated using

$$F'(x,y) = \frac{2}{N}\sum_{u=0}^{N-1}\sum_{v=0}^{N-1} C(u)C(v)\, D(u,v) \times \left[\cos\left(\frac{\pi u(2x+1)}{2N}\right)\right]\left[\cos\left(\frac{\pi v(2y+1)}{2N}\right)\right] \quad (5)$$

$$C(v), C(u) = \begin{cases} \frac{1}{\sqrt{2}} & if\ u=0 \\ 1 & otherwise \end{cases} \quad (5.1)$$

Combine all $8 \times 8$ windows to produce the stego-image. This stego-image along with the secret key is sent as in [17].

### 3.1.1 OPTIMIZING CAPACITY AND STEGO-IMAGE QUALITY USING ADAPTIVE K

The scheme suggested in [17], [21] divide image in $N \times N$ window size where $N$ varies from 8 to 256. Increasing window size results in achieving higher payload capacity that decreases stego-image quality. Every steganography algorithm needs such flexibility for practical applications. One of the application scenario is discussed in [17], [21], where we need to find an optimal cover image from a set of host images.

Our scheme is different from the above. Although we keep the window size fixed ($8 \times 8$), but we adaptively change the matrix size of dominant cofficients ($K \times K$) in the left top corner of every window (as shown in Fig. 2d). This gives us optimal combination of payload capacity and imperceptibility. However, in order not to let the stego-image imperceptibility fall below a certain level, $K_{min}$ is chosen so that $K \geq K_{min}$.

### 3.2 RETRIEVING PROCESS

The retrieving process comprises of six steps that takes stego image to generate a secret image. These steps are explained below,

**STEP 1:** The stego-image acquired along with the key is initially splitted into $8 \times 8$ window and then each window is transformed to its frequency domain equivalent using (2).

**STEP2:** we have total 64 coefficients in a $8 \times 8$ DCT coefficient window. Furthermore, the size of host image Discrete Cosine Transform coefficients square block in the top left corner of each window is also known. Rest of the coefficients belong to the secret message. These coefficients are obtained.

**STEP3:** The retrieve message in last step is a normalized secret image. This image is readjusted by using $NF$ present in secret key by using,

$$SI = \left[\frac{NSI}{NF}\right] \times 255 \quad (6)$$

and rearranged to generate the secret image $SI$.

### 4. EXPERIMENTAL COMPARISON

This section analyses the result of the proposed scheme with comparison to the recently proposed schemes. The dataset used to examine suggested algorithm is shown in Fig. 3. This dataset is exactly same as used in [17], [21], [24]. This data set spans various color structure details. In [17], the author used different window size varying from $8 \times 8$ to $256 \times 256$. This algorithm has a fix window size that is $8 \times 8$. The constraint $K_{min}$ already discussed in the previous section can limit the embedding capacity for improving PSNR. Finally, this scheme is less complex in term of computations for the scenario proposed in [17].

#### 4.1 RESULTS COMPARISON

Secret image (flower image) is injected in five different cover images (peppers, snow tiger, balloons, zebras and tomatoes) shown in Fig. 3. The scheme proposed in [17] achieved the best payload capacity with the "tomatoes" image that was 20.31 bpp and 26dB PSNR. Another scheme proposed in [25], achieved the capacity of 15.1 bpp with the 18.4 dB PSNR. Author of [21] proposed a scheme based on discrete cosine transform to achieve embedding capacity of 20.22 bpp with 25.1 dB PSNR. The method proposed by Lee & Chen in [8] embeds a secret message with the payload capacity of 12.18 bit per pixel with 34.03 dB PSNR. The scheme proposed in [9] achieved the PSNR up to 40db that was relatively higher than other proposed scheme but the payload capacity was only 6 bpp.

However, the method proposed in this work embeds the secret message at 21.5 bpp with a 38.24 dB PSNR in "tomatoes" cover image. These results are achieved with $adaptive\ k$ i.e. no constraint on embedding limits. If the constraint "$K_{min} = 4$" it means that at least block size $4 \times 4$ of host image discrete cosine transform coefficients must be retained in each $8 \times 8$ window of stego-image. This constraint $K_{min} = 4$ results in payload capacity of 13.37 bit per pixel with 44.7 dB PSNR in the similar host image. Table 1 represents the comparison of suggested schemes with the already existing schemes.

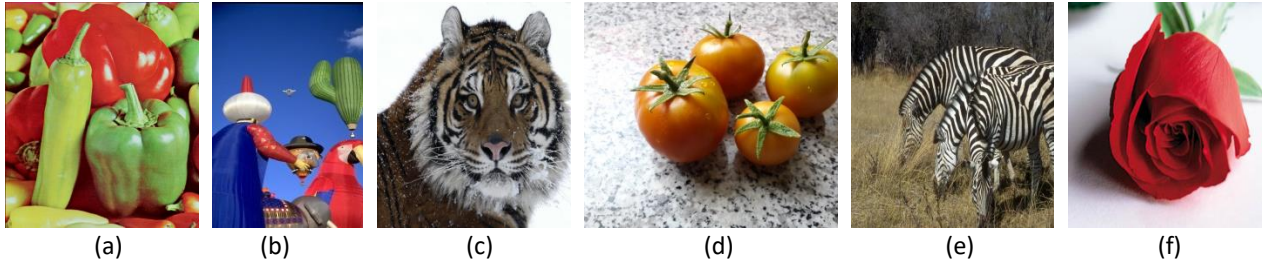

(a)     (b)     (c)     (d)     (e)     (f)

**Fig.3** a to e cover images. f secret image.

**Table 1.** comparison in term of payload capacity and Peak Signal to Noise Ratio (PSNR).

| Algorithm | Payload Capacity (bpp) | PSNR (dB) |
|---|---|---|
| Brisbane et al.[9] | 6 | 40 |
| Lee & Chen [8] | 12.18 | 34.03 |
| Saeed & Shahrokh [25] | 15.1 | 18.4 |
| Rabie & Kamel [21] | 20.22 | 25.1 |
| GAR-DCT(256x256) [17] | 20.31 | 26 |
| Proposed Scheme ($K_{min} = 4$) | 13.37 | 45.22 |
| Proposed scheme ($adaptive\ K$) | 21.5 | 38.24 |

**Table 2.** Comparison between GAR-DCT proposed in [16] and the proposed algorithm.

| Host Image | GAR DCT | | | Proposed scheme (window size $8 \times 8$) | | |
|---|---|---|---|---|---|---|
| | Window size | Capacity (bpp) | PSNR (dB) | Constraint | Capacity (bpp) | PSNR (dB) |
| Pepper | $8 \times 8$ | 10.15 | 32.8 | $K_{min} = 4$ | 16.45 | 32.88 |
| Balloons | | 11.09 | 36.4 | | 16.45 | 42.9 |
| Snow Tiger | | 9.03 | 27.9 | | 16.45 | 37.94 |
| Tomatoes | | 8.1 | 37.6 | | 16.45 | 43.05 |
| Zebras | | 7.81 | 32.5 | | 16.44 | 37.06 |
| Pepper | $64 \times 64$ | 17.64 | 28 | $K_{min} = 3$ | 18.84 | 32.29 |
| Balloon | | 17.25 | 32.2 | | 18.84 | 39.34 |
| Snow Tiger | | 17.25 | 24.9 | | 18.76 | 37.39 |
| Tomatoes | | 16.62 | 29.4 | | 18.85 | 41.87 |
| Zebras | | 17.17 | 23.0 | | 18.69 | 34.87 |
| Pepper | $128 \times 128$ | 18.21 | 27.4 | $K_{min} = 2$ | 20.43 | 29.53 |
| Balloons | | 18.21 | 31 | | 20.48 | 33.69 |
| Snow Tiger | | 18.79 | 24.3 | | 20.26 | 34.62 |
| Tomatoes | | 18.29 | 27.8 | | 20.55 | 40.2 |
| Zebras | | 19.88 | 18.6 | | 20.07 | 34.19 |
| Pepper | $256 \times 256$ | 18.13 | 27.9 | $adaptive\ K$ | 21.14 | 28.25 |
| Balloons | | 19.2 | 28.8 | | 21.34 | 31.59 |
| Snow Tiger | | 20.48 | 22.8 | | 20.07 | 32.05 |
| Tomatoes | | 20.31 | 26 | | 21.5 | 38.24 |
| Zebras | | 20.05 | 18.9 | | 20.81 | 31.48 |

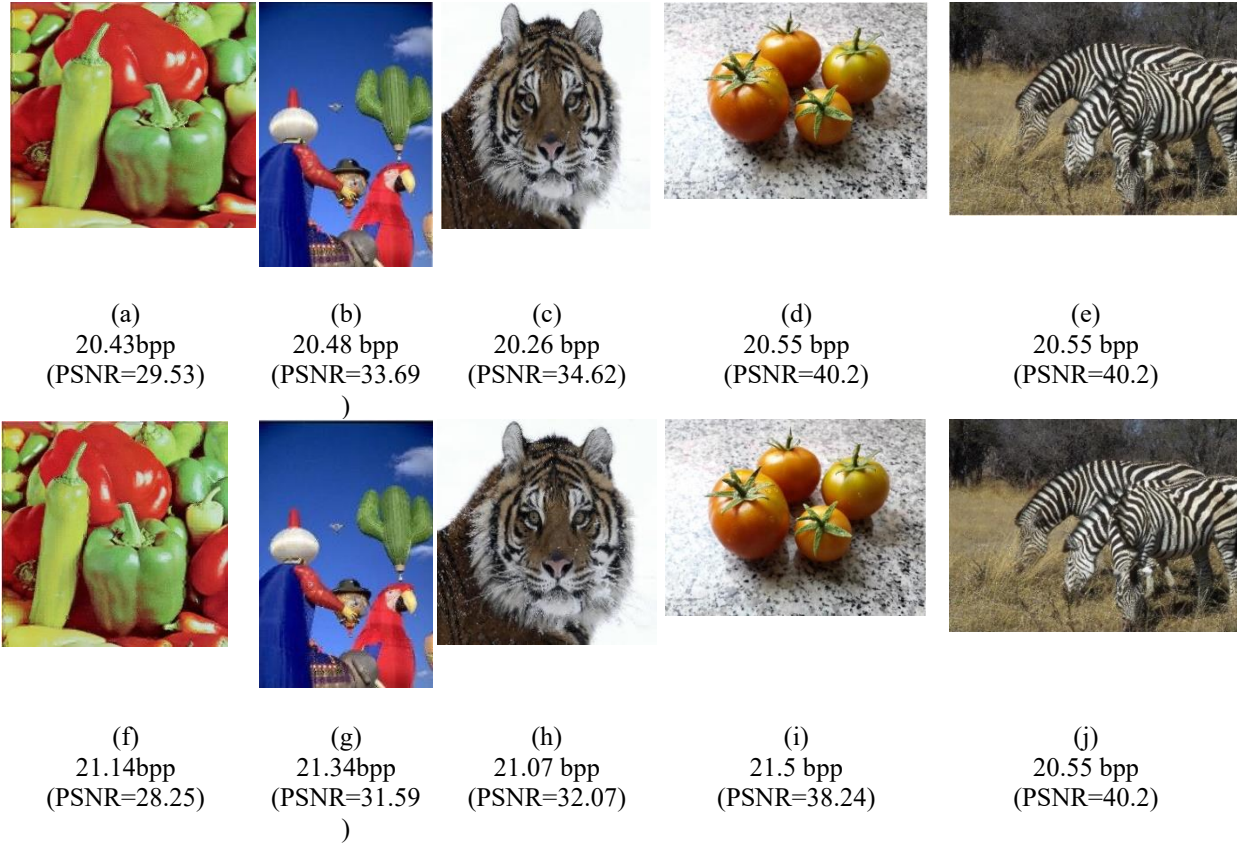

| (a) | (b) | (c) | (d) | (e) |
| --- | --- | --- | --- | --- |
| 20.43bpp | 20.48 bpp | 20.26 bpp | 20.55 bpp | 20.55 bpp |
| (PSNR=29.53) | (PSNR=33.69) | (PSNR=34.62) | (PSNR=40.2) | (PSNR=40.2) |

| (f) | (g) | (h) | (i) | (j) |
| --- | --- | --- | --- | --- |
| 21.14bpp | 21.34bpp | 21.07 bpp | 21.5 bpp | 20.55 bpp |
| (PSNR=28.25) | (PSNR=31.59) | (PSNR=32.07) | (PSNR=38.24) | (PSNR=40.2) |

**Fig.4** Stego-images, a to e "$K_{min}=2$". f to j *adaptive k*

### 4.2 PROPOSED SCHEME RESULTS

Result obtained by the suggested algorithm on the cover images shown in Fig. 3. The flower image is injected as a secret message is shown in Fig. 3. Fig. 4 shows different stego-images after embedding the flower image. Instead of taking different window size as in [17], [21] the suggested algorithm takes a fixed window size of $8 \times 8$ but by putting constraint "$K_{min}$" over the maximum embedding limits improves the PSNR. Results in Table 2 obviously show that the suggested algorithm has achieved considerable improvements in term of peak signal to noise ratio and payload capacity.

### 4.3 EXAMPLE SCENARIO

This algorithm can be used for the scenario proposed in [17], [21] where one has to select an image that yields us better PSNR on a desired payload capacity. As an example, suppose if the capacity of 20 bit per pixel is needed for the chosen data set. Tomatoes image with constraint "$K_{min} = 2$" is the best option that results in PSNR up to 40 db.

### 5. COMPUTATIONAL COMPLEXITY

The schemes proposed in [17], [21] takes an image and transform using discrete cosine transform on multiple window sizes i.e. $8 \times 8$, $64 \times 64$, $128 \times 128$ $and$ $256 \times 256$ and then embeds $SI$ and calculates the embedding capacity and PSNR. Unlike the previous, proposed algorithm takes 2D-DCT of $8 \times 8$ block only and can still achieve superior results with regard to payload capacity and stego-image quality. Hence the scheme suggested in this work requires fewer computations for the scenario discussed in section 4.3. Furthermore, the concept of using $k_{min}$ constraint on embedding limits gives us flexibility to improve imperceptibility at the cost of capacity.

| cover image size | Window size | Total windows | Additions per window[26] | Multiplication per window[26] | Total Additions | Total multiplications |
|---|---|---|---|---|---|---|
| 512 × 512 | 64 × 64 | 8 | 62442 | 13528 | 499536 | 108224 |
| 512 × 512 | 128 × 128 | 4 | 291434 | 62552 | 1165736 | 250208 |
| 512 × 512 | 256 × 256 | 2 | 1331050 | 283480 | 2662100 | 566960 |
| Total number of additions and multiplications for 64 × 64, 128 × 128, 256 × 256 window size. | | | | | 4327372 | 925392 |

The required multiplications and additions for calculating 2D-DCT of $N \times N$ block using an algorithm proposed in[26], are given by

$$\frac{N^2}{2}\log_2 N + \frac{N^2}{3} - 2N + \frac{8}{3} \tag{7}$$

$$\frac{5N^2}{2}\log_2 N + \frac{N^2}{3} - 6N + \frac{62}{3} \tag{8}$$

Using the above mentioned equations we conclude that the proposed algorithm requires 925392 multiplications and 4327372 additions less than the
algorithm proposed in [17], [21] for computing 2D-DCT. Extra cost for calculating 2D-DCT of window size $64 \times 64, 128 \times 128$ and $256 \times 256$ on an image size $512 \times 512$ is shown in Table 3.

## 6. CONCLUSION

This work presents a discrete cosine transform based steganography slgorithm that attain improved results in respect of payload capacity and PSNR than already proposed algorithms. The concept is to retain a square block of cover image discrete cosine transform coefficients in the top left corner of a cover image. The rest of the space in a $8 \times 8$ discrete cosine transform block is replaced with the secret message normalized intensity values to generate a stego-image. Embedding data in a non-square block give us improved payload capacity. Furthermore, the normalization factor for each window is chosen according to the statistics of that window resulting in improved stego-image quality. The concept of $k_{min}$ is introduced as a constraint on maximum embedding capacity and gives us a tendency to play between payload capacity and stego-image quality. Finally, the application of the proposed scheme is demonstrated by a scenario where we have to find an optimal cover image that gives us a best stego-image quality on a required payload capacity.